%% file: main-arxiv.tex
\documentclass[lettersize,conference]{IEEEtran}
\IEEEoverridecommandlockouts

\usepackage[numbers]{natbib}  
\usepackage{hyperref}
\usepackage{listings}
\usepackage{graphicx}
\usepackage{textcomp}
\usepackage{xcolor}
 \usepackage{url} 
 \usepackage{makecell}
 \usepackage{mdframed}
 \usepackage{tabularx}
 \usepackage{booktabs} 
 \usepackage{adjustbox}
\usepackage{tablefootnote} 
 \newmdenv[
  backgroundcolor=blue!5,
  hidealllines=true,
  roundcorner=5pt,
  innerleftmargin=10pt,
  innerrightmargin=10pt,
  innertopmargin=10pt,
  innerbottommargin=10pt,
  font=\small,
  skipabove=10pt 
]{keyfindingsbox}

\usepackage[table]{xcolor}
\usepackage{listings}

\definecolor{diffadd}{RGB}{0,128,0}
\definecolor{diffdel}{RGB}{180,0,0}
\definecolor{diffctx}{RGB}{60,60,60}
\definecolor{diffheader}{RGB}{0,0,150}

\lstdefinelanguage{diff}{
  morecomment=[f][\color{diffheader}]{diff},
  morecomment=[f][\color{diffheader}]{+++},
  morecomment=[f][\color{diffheader}]{---},
  morecomment=[f][\color{diffheader}]{@@},
  morecomment=[f][\color{diffadd}]{+},
  morecomment=[f][\color{diffdel}]{-},
}

\definecolor{listingbg}{RGB}{245,242,236}

\lstdefinelanguage{diff}{
  morecomment=[f][\color{gray}]@@,
  morecomment=[f][\color{red}]-,
  morecomment=[f][\color{teal}]+,
}

\usepackage{multicol}
\usepackage[most]{tcolorbox}
\usepackage{listings}
\usepackage{xcolor}

\definecolor{codegreen}{rgb}{0.0, 0.6, 0.0}
\definecolor{codered}{rgb}{0.8, 0.0, 0.0} 

\definecolor{codebg}{rgb}{0.97,0.97,0.97}  

\lstdefinestyle{code}{
  basicstyle=\ttfamily\footnotesize,
  keywordstyle=\color{blue},
  commentstyle=\color{gray},
  stringstyle=\color{red},
  breaklines=true,
  columns=fullflexible,
  keepspaces=true,
  showstringspaces=false,
  backgroundcolor=\color{codebg},
  frame=none,
  escapeinside={(*@}{@*)}
}

\def\BibTeX{{\rm B\kern-.05em{\sc i\kern-.025em b}\kern-.08em
    T\kern-.1667em\lower.7ex\hbox{E}\kern-.125emX}}
\begin{document}


\title{Patterns of Developer Adoption of LLM-Generated Code Refactoring Suggestions}

\author{  
  David Schön$^1$, Faiza Amjad$^1$, Tehreem Asif$^1$,\\ Ranim Khojah$^1$, Mazen Mohamad$^{1,2}$, Francisco Gomes de Oliveira Neto$^1$, Philipp Leitner$^1$\\
  $^1$\textit{Chalmers University of Technology and University of Gothenburg}, $^2$\textit{RISE Research Institutes of Sweden}\\                                  
  Gothenburg, Sweden \\
  khojah@chalmers.se, francisco.gomes@cse.gu.se, mazen.mohamad@ri.se, philipp.leitner@chalmers.se}            
 \maketitle

\begin{abstract}
Large language models (LLMs) have gained widespread popularity and have steadily improved over time, enabling software developers to use them for various code-related tasks. One common task is code refactoring, where the LLM suggests changes for the developer to apply to their code to improve quality attributes such as readability or maintainability. While current research focuses on evaluating LLM-generated refactoring suggestions, there is a limited understanding of how developers apply these suggestions in practice. To explore this, we analyze 169 GitHub commits where developers refactor their code based on a ChatGPT conversation linked in the commit message. We found that developers mostly accept and use the suggestions without modifications. When changes are made, they are mostly major and fall into five different patterns that depend on the refactoring activity, the developer's prompt, and the validity of the response from ChatGPT.
%

\end{abstract}


\begin{IEEEkeywords}
Large language models, Code refactoring, AI-assisted development
\end{IEEEkeywords}




\input{introduction}
\input{background}
\input{methodology}

\input{results}
\input{discussion}

\input{conclusion}

\section*{Acknowledgement}
The paper is based on a thesis project done by the first three authors.
This work was partially supported by the Wallenberg AI, Autonomous Systems and Software Program (WASP) funded by the Knut and Alice Wallenberg Foundation.

\bibliographystyle{ieeetr}
\bibliography{bib}


\end{document}

%% file: introduction.tex
\section{Introduction}
\label{sec:Intro}
In recent years, software development has increased in complexity and requires constant effort to improve code quality and maintainability \citep{1265817}. One common practice to attain this is code refactoring. Refactoring is defined as improving the internal structure of code \textit{without changing its external functionality} with the goal of improving various aspects of the code quality \citep{kaur2016analysis}.
At the same time, rapid advancements in Artificial Intelligence (AI) have led to the emergence of Large Language Models (LLMs), which are increasingly being integrated into the software development workflow \citep{githubreport} due to their ability to generate code \citep{10403378}. 

Refactoring can be time-consuming or prone to errors, but LLMs offer different ways to support developers in refactoring activities, such as providing a refactored version of their code \cite{cordeiro2024empirical}. While LLMs have shown to provide effective refactoring suggestions \citep{metsola2024large}, it remains unclear how developers actually use and integrate those suggestions in their code base. 

Recent research indicates that modifications during the validation of LLM-generated refactoring are often required \cite{liu2024empiricalstudypotentialllms}.
While some developers might directly implement suggestions from LLMs, others may need to make modifications before using them. Therefore, understanding these modifications and the developer's behavior is important to improve LLM-based chatbots and AI-assisted development workflows. 

This paper investigates how developers use LLM-generated refactoring suggestions, more specifically GPT-models, through ChatGPT. 
This includes investigating what parts of the suggestions that they decide to keep, change or remove before applying to the source code.
Unlike previous research, we go beyond the refactoring outcome of the LLM and study the step where the developer adopts the suggestion during software development or maintenance. We achieve this by utilizing lexical similarity measures to automatically estimate how much of the generated refactoring suggestion ended up being used in the final refactored code, as well as a qualitative manual analysis of the types of modification that developers often make on the suggestions before applying them.
Hence, we formulate our research questions as follows: \\

\noindent\textbf{RQ1. To what extent do developers adopt ChatGPT-generated code refactoring suggestions in practice?} 

We conduct a quantitative analysis on a dataset that we construct based on the DevGPT dataset \cite{DevGPT_Dataset}. We use three lexical similarity metrics to measure the code similarity between refactoring suggestions in a ChatGPT conversations and the corresponding files refactored in a commit. Specifically, we use the rate of token match, Jaccard n-gram similarity and normalized Levenshtein similarity. We found that the majority of the developers adopted ChatGPT refactoring suggestions with either no changes (very high similarity) or with major changes (low similarity). \\
    
\noindent\textbf{RQ2. What changes do developers typically make to ChatGPT-generated refactoring suggestions before committing?}
    
We perform a qualitative analysis by manually inspecting a subset of code suggestions and their respective implementations in the commits. This allows us to better understand the type of adjustments made by the developers and identify patterns in the generated suggestions. One finding is that developers fully adopt refactoring suggestions related to utility functions (e.g., renaming and reorganizing files), while more complex suggestions often cause errors or add unwanted behavior to the code, which requires developers to make more substantial modifications on the suggestion.


    
    


Our scientific contributions are: (i) an empirical study on how developers use LLM-generated refactoring suggestions in practice, going beyond prior work that focused solely on evaluating the suggestions; and (ii) a method to estimate adoption levels of these suggestions using similarity metrics. On the practical side, our findings on how developers modify suggestions provide insights into when LLMs can be automated versus used collaboratively during refactoring in LLM-assisted development processes.
Moreover, we share our curated dataset with a replication package, aiming to support future studies in the area.



%% file: background.tex
\section{DevGPT Dataset}
\label{sec:devgpt}

DevGPT is an open-source repository that is used to study the interaction of developers with large language models, specifically ChatGPT. The data set from \citet{DevGPT_Dataset} contains 29,778 prompts and responses from ChatGPT conversations (19,106 code snippets) linked with corresponding software development artifacts, such as source code, commits, change logs, issues, etc. The data was collected between July 2023 and August 2023, during which ChatGPT models GPT-3.5 and GPT-4 were used.

This data set consists of a total of 3245 commit objects for evaluation, which provides valuable information on how developers engage with ChatGPT-generated code suggestions, either in the form of written instructions or code snippets. Given the scale, DevGPT serves as a strong foundation for analyzing the extent to which developers adopt, modify, or reject ChatGPT's refactoring recommendations. 
To connect these changes back to the original ChatGPT suggestions, we introduce the use of Levenshtein similarity to map suggestions to the most relevant changed files. We then measure the similarity between the mapped files and their corresponding suggestions to estimate the degree of adoption. Finally, we manually categorize the types of modifications developers made to the original suggestions.

\section{Related Work}

Recent research has been focusing on exploring the capabilities of LLMs in supporting software engineers in their tasks \cite{khojah2024beyond,guo2023exploringpotentialchatgptautomated} and how developers interact with these LLMs \cite{khojahFromHuman,brynjolfsson2024generativeaiwork}. 
Code refactoring is an important software engineering practice that aims to improve the design and structure of existing code, e.g., to enhance its readability and maintainability, while preserving its behavior \cite{fowler2018refactoring}. Hence, LLM's ability to improve code quality through refactoring has been studied and empirically shown to even out-perform human developers when it comes to certain quality aspects, such as reducing code smells \cite{cordeiro2024empirical}. 
Collaborating with LLMs for code refactoring tasks has also been studied by exploring refactoring-related conversations between human developers and ChatGPT to identify prompt patterns \cite{paper2} as well as capturing how the developers identify areas of improvement in the code and how the LLMs respond \cite{paper1}. Moreover, researchers have evaluated the effectiveness of refactoring suggestions by analyzing the resulting code quality and show that LLMs can refactor code effectively but it costs readability as it becomes more concise and compact making it harder for developers to understand \cite{metsola2024large}.

Despite their capabilities and potential in supporting code refactoring tasks, LLMs still have limitations and cannot be considered fully reliable as software development tools. For example, one of the most commonly used LLMs, ChatGPT, is found to be equally confident when generating correct code as it does when generating incorrect code \cite{Csuvik2023}. Additionally, ChatGPT-generated code can suffer from code quality issues \cite{liu2023refiningchatgptgeneratedcodecharacterizing}. For refactoring tasks, the LLMs might in some cases introduce unsafe changes \cite{liu2024empiricalstudypotentialllms}. The non-deterministic nature of LLMs can also lead to different outputs even when provided the same prompt, which threatens their reliability \cite{GPTCapabilities}.
This leads to the question whether the LLM output can be directly used in practice by developers, e.g., for code refactoring. 

In contrast to the related work, this study gains insights into how the code refactoring suggestions are integrated in the code in practice. This is done by analyzing the refactored code along with the LLM suggestions to determine the extent to which these suggestions have been adopted.

%% file: methodology.tex
\section{Methodology}
\label{sec:method}
\begin{figure*}[!ht]
    \centering
    \includegraphics[width=0.9\linewidth]{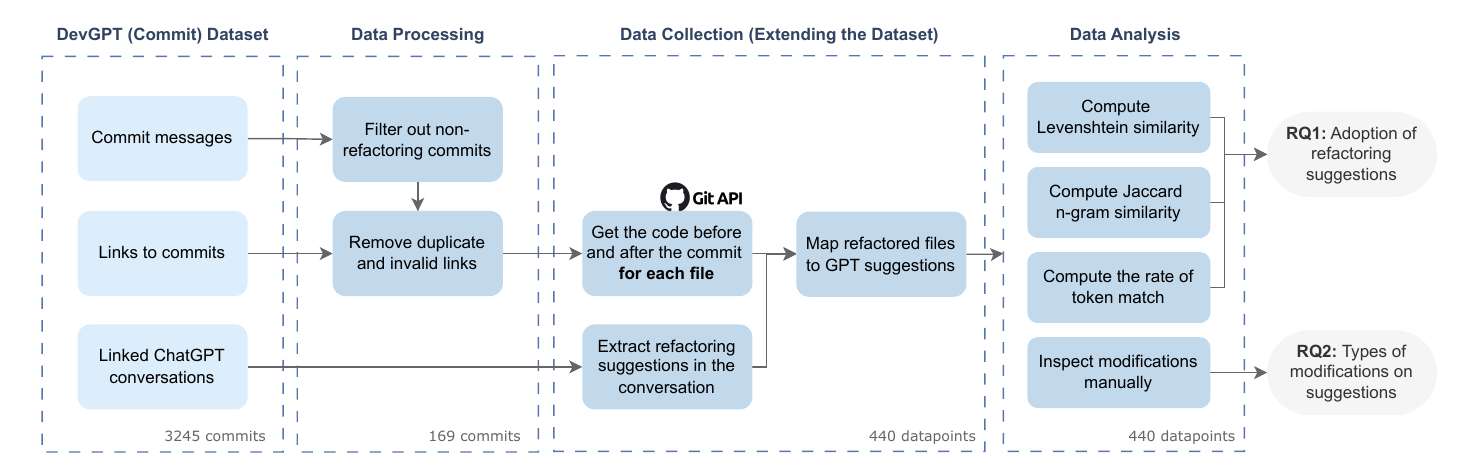}
    \caption{Overview of the process we followed in our study.}
    \label{fig:method}
\end{figure*}

To understand how developers implement GPT-generated refactoring suggestions, we follow the process we present in Figure \ref{fig:method}. On a high level, we pre-process the DevGPT dataset (see Section \ref{sec:devgpt}) by removing duplicate commits and filtering out ones that are not related to a refactoring activity. Then we extend the dataset by extracting the files changed in the commit on GitHub as well as the different code blocks that were suggested in the ChatGPT chat linked in the commit message. Finally, we map each file in a commit to a refactored code that is suggested by ChatGPT. In our analysis, we quantitatively compute different similarity measures between the refactoring suggestions and the final code that the developer committed after applying the suggestion. We also manually inspect a subset of the dataset to understand the types of modification that developers make on LLM-generated suggestions before integrating them into their code base.

\subsection{Data processing} 
\label{sec:processing}
To be able to compare the generated refactoring suggestions and the code changes implemented in the commit, we construct a dataset based on DevGPT that includes 440 files that belong to 169 commits along with the corresponding suggestion retrieved from a ChatGPT conversation linked in the commit message. The dataset and the scripts used to build it are available in our replication package \cite{replication}.

We first compile a list of refactoring-related keywords from literature. In particular, we select specific keywords that describe refactoring activities \citep{Ratzinger2007} and operations to improve the code design \citep{1265817}. We also note that previous research showed that developers often use generic terms, such as ``refactor'' or ``optimize'' to describe their refactoring task and do not necessarily specify the type of refactoring activity they performed \citep{alomar2021documentationrefactoringtypes,6112738}.

To ensure having a complete list of keywords, we extract all keywords that exist in all commit messages in the DevGPT dataset. We tokenize and lemmatize the commit messages, remove punctuations and stop words, then store them in a set to remove duplicates. This results in a list of 2918 words used in all commit messages. Two authors collaboratively review these keywords over five sessions and select keywords relevant to code refactoring guided by the keywords extracted from literature. This collaborative process intends to mitigate individual bias and ensure consistency in the selection.

Finally, we merge both lists and then add possible synonyms and British/American spelling variants when applicable, resulting in 25 refactoring keywords in Table \ref{tab:refactor-keywords}. Our aim is to make the list of keywords usable by future research by including keywords that may not appear in the DevGPT dataset specifically but can be used in other refactoring-related commit messages.

\begin{table}[!ht]
\footnotesize
\centering
\begin{center}
\caption{Final list of refactoring keywords.}
\label{tab:refactor-keywords}
\begin{tabular}{ c c c c c }
Refactor & Rewrite & Simplify & Organize & Reorganize \\
Rename & Regenerate & Restructure & Reformat & Clean \\
Duplicate & Improve & Optimize & Redundant & Split \\
Unify & Unused & Useless & Enhance & Complexity \\
Refine & Polish & Organise & Reorganise & Optimise \\
\end{tabular}
\end{center}
\end{table}

Finally, we use the refactoring keywords to filter the commits in the dataset, then we remove redundant commits and ones that link to an invalid GitHub URL. There is a risk that the automated filtering leads to false positives in the data (we discuss this further in our threats to validity).
The filtration process resulted in a curated dataset of 169 commits.

\subsection{Data collection}
\label{sec:collection}
After processing the dataset of 169 refactoring commits, we extend it by adding two additional fields to each commit object, namely a list of refactoring suggestion objects and a list of changed files objects.
We extract the code blocks from the GPT-generated answers in the conversation and assign them unique IDs. The dataset also includes meta-data, including the programming language used and the GPT model that was used to generate the answers.

Then, we use the GitHub API\footnote{\url{https://docs.github.com/en/rest}} to retrieve information from each commit, such as the changed files. The response contains a ``unified diff'' of changes for each file. The unified diff is a collection of lines. Each line is separated by a newline character followed by: ``\texttt{-}'', ``\texttt{+}'', or `` '', to indicate deletion, addition, or no change, respectively. We use this to reconstruct two variations of the files, that is, before implementing the GPT-generated refactoring suggestion (i.e., \texttt{before\_changes}), and after applying the suggestion (i.e., \texttt{after\_changes}). In Listing \ref{lst:unified-diff}, we show how a unified diff looks like; we create the \texttt{before\_changes} by adding the unchanged and deleted lines (red), and the \texttt{after\_changes} file by adding the unchanged lines and the added lines (green). We also store the added lines separately for further analysis.


\begin{lstlisting}[
style=code,
  frame=none,
  rulecolor=\color{white},
  backgroundcolor=\color{codebg},
  caption={Visualization of a unified diff including added (green), deleted (red) and unchanged lines (default).},
  label={lst:unified-diff}
]
diff --git a/src/frontend/components/RequirementsEditor.jsx b/src/frontend/components/RequirementsEditor.jsx
index 0000000..0000000 100644
--- a/src/frontend/components/RequirementsEditor.jsx
+++ b/src/frontend/components/RequirementsEditor.jsx
@@ -1,32 +1,26 @@
 import { createEffect } from 'solid-js';
-import { requirements, setRequirements } from '../model/requirements'; // Imported setRequirements
+import { promptDescriptor, setPromptDescriptor } from '../model/promptDescriptor'; // Added setPromptDescriptor
 import { getYamlEntry } from '../service/getYamlEntry';
+import jsyaml from 'js-yaml'; // Importing the YAML parser
 
-let lastPostedTime = 0;
-let lastThrottledValue = null;
-
 const RequirementsEditor = () => {
   const handleRequirementsChange = async (e) => {
-    const now = Date.now();
-    if (now - lastPostedTime < 1000) {
 ...
\end{lstlisting}

We exclude certain file extensions and directories that could introduce noise into our dataset. This includes files inside common dependency, build, and environment folders (e.g., \texttt{site-packages/} or \texttt{node\_modules/}), as well as files with extensions that usually do not contain source code (e.g., \texttt{.gitignore} or \texttt{.yaml}). The complete list of excluded file extensions and directories can be found in our replication package \citep{replication}.

After collecting all relevant fields, we map each changed file to a code block shared in the ChatGPT conversation (i.e., the refactoring suggestion). This mapping is based on normalized Levenshtein similarity \citep{yujian2007normalized}. Specifically, for each \texttt{after\_changes} file, we calculate its similarity to every code suggestion in the corresponding ChatGPT conversation, then assign it to the code suggestion with the highest similarity score.

The mapping step resulted in a dataset of 440 datapoints. Each datapoint includes a unique refactoring\_ID, a file before refactoring, the refactoring suggestion, the file after refactoring, and additional information related to the commit message and ChatGPT conversation. We present an overview of the dataset in Table \ref{tab:repo_stats}. We noticed that one project is significantly larger than the others in terms of commits and files modified. While this introduces some imbalance, we prioritize covering as many refactoring scenarios as possible, regardless of the size of the repository they belong to.

\begin{table}[!ht]
\centering
\scriptsize
\caption{Overview of analyzed data: number of commits, files, prompts, and per-commit averages with standard deviations.}
\begin{tabularx}{\columnwidth}{p{2.1cm}p{0.8cm}p{0.3cm}p{0.7cm}ll}
\toprule
\textbf{Repo Name} & \textbf{Commits} & \textbf{Files} & \textbf{Prompts} & \textbf{Avg Files} & \textbf{Avg Prompts} \\
\midrule
tisztamo\-/Junior                       & 143 & 407 & 238 & 2.8 ($\pm$ 1.8) & 1.7 ($\pm$ 1.1) \\
pbrudny\-/jobsforit-de                  & 7   & 10  & 48  & 1.4 ($\pm$ 0.8) & 6.9 ($\pm$ 5.1) \\
hoshotakamoto\-/banzukesurfing          & 6   & 17  & 38  & 2.8 ($\pm$ 1.0) & 6.3 ($\pm$ 5.1) \\
REReal8\-/CCWorldPlatform                & 1   & 2   & 20  & 2.0 (NA) & 20.0 (NA) \\ 
openai\-/evals                          & 1   & 1   & 14  & 1.0 (NA) & 14.0 (NA) \\
labd\-/terraform-provider-storyblok     & 1   & 2   & 9   & 2.0 (NA) & 9.0 (NA) \\
SupraSensum\-/TOP-project-etch-a-sketch & 1   & 1   & 1   & 1.0 (NA) & 1.0 (NA) \\
\midrule                            
Total                               & 160 & 440 & 368 & 13.1 & 58.9  \\
\bottomrule
\end{tabularx}

\label{tab:repo_stats}
\end{table}

\subsection{Data analysis}
\label{sec:analysis}

We analyze how developers adopted 440 refactoring suggestions in their code using both a quantitative approach and a manual qualitative approach.


\subsubsection{Quantitative analysis}

We focus on different lexical-based similarity metrics, that is, normalized Levenshtein similarity \citep{yujian2007normalized}, Jaccard n-gram similarity \citep{jaccard1901etude}, and the rate of matched tokens which we introduce later in this section. 
We found lexical-based metrics to be more suitable for capturing surface-level edits that are commonly found in refactoring (e.g., renaming variables or re-structuring code blocks). We also complement the analysis with CrystalBLEU \cite{eghbali2023crystalbleu} to capture semantic-level modifications (e.g., structural refactoring, control-flow changes).



We compare the similarity between ``after\_changes'' file and the corresponding ChatGPT suggestion, rather than limiting the comparison to only the changed lines. To capture edits such as line deletions, that would otherwise be missed, we used two similarity metrics. We measure Jaccard n-gram similarity, which compares the sets of overlapping n-grams from two code snippets. Specifically, we use 3-grams, as many programming language keywords and common identifiers are around three characters long (e.g., \texttt{for}, \texttt{int}, \texttt{new}). This makes 3-grams well-suited for capturing patterns and re-orderings in code.
 
To complement Jaccard, we use the Levenshtein distance which is based on the edit distance between two strings, that is, the minimum number of single-character insertions, deletions, or substitutions required to transform one string into the other. We use the normalized Levenshtein similarity, which scales this distance by the length of the strings to produce a similarity score between 0 and 1, where higher values indicate greater similarity.
This helps us to identify minor modifications in each line of code, such as modifications of variable names or other small tweaks by the developer.

However, during the analysis, we found that many refactoring suggestions involved changes to multiple files. In these cases, the model often provided a single large code block that included code for all affected files. While we correctly mapped this suggestion to each relevant file, the similarity scores were artificially low. This is because only part of the suggestion matched the code in each file, making it appear less similar even if the actual match was high.

To address this, we introduce a new metric (Token Match Rate) which measures the proportion of tokens in the committed refactored file that also appear in the refactoring suggestion.

{\small
\[Token Match Rate = \frac{Number Of Matched Tokens }{Total Tokens In Committed File}\]}


All three measures are meant to complement each other in capturing different aspects of similarity between the refactoring suggestion and the final refactored code. For example, the token match rate can handle complex suggestions but does not capture deletions, which in turn can be captured by Jaccard and Levenshtein.

We were also interested in measuring semantic similarity using CrystalBLEU to assess whether it offers additional insight into the modifications made to the suggestion prior to its application. CrystalBLEU captures code-specific features and filters out common and uninformative n-grams (e.g., for, if) to focus on meaningful changes on the code.

\subsubsection{Qualitative analysis}
\label{sec:qualitative}

To understand the behavior of developers when interacting with ChatGPT refactoring suggestions, we manually inspect what parts of the ChatGPT-suggested code they keep, change or remove in a subset of 190 datapoints. 
We selected the subset based on the token match rate results, where we included all datapoints that indicated a modification on the suggestion (similarity lower than 1.0), the total of these datapoints were 96. We also selected 94 with token match rate exactly 1.0.

In the manual inspection, we look at the committed file before refactoring, the refactoring suggestion in the ChatGPT conversation, and the refactored version of the code. We focus on understanding the i) the refactoring steps that were suggested by ChatGPT, and the ii) types of modification that the developer make on the suggestion before implementing it. During the analysis, we observed that developers can use the term refactoring in commits that change the behavior of the code. In particular, we noticed in 16 cases that the commit message explicitly mentions refactoring, but the changes concern implementing edge-case or error handling, or adding more elements to the frontend (e.g., a navigation bar). Since these activities change the behavior of the code, we excluded them from our qualitative analysis.

The inspection was carried out by two of the authors at the same time in two collaborative sessions. This approach allowed us to limit individual bias and discuss disagreements to resolve them on the spot.

%% file: results.tex
\section{Results}
\label{sec:results}

This section presents the findings obtained from quantitatively and qualitatively analyzing our dataset that we constructed from the DevGPT dataset. We present the results of comparing the code suggestions from ChatGPT with the final commit implementations to answer our research questions about the adoption of ChatGPT refactoring-related suggestions. 

\begin{figure}[!ht]
    \centering
    \includegraphics[width=\linewidth]{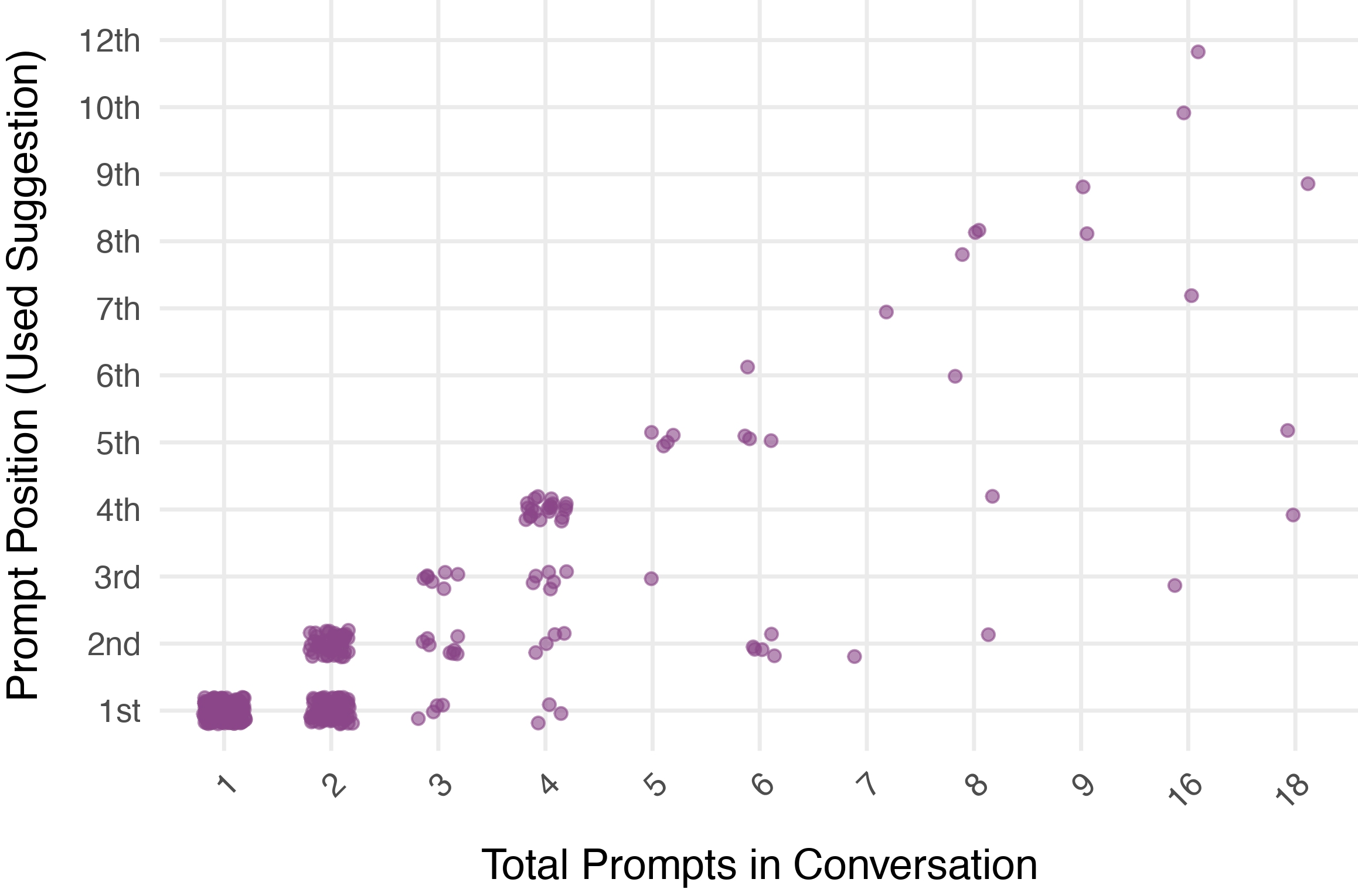}
    \caption{Relative position of the adopted prompt (e.g., 1st, 2nd, etc.) in the conversation (y-axis) vs. the total number of prompts in 440 conversations (x-axis).}
    \label{fig:adopted-prompts}
    \vspace{-5pt}
\end{figure}

\begin{figure*}[!ht]
    \centering
    \includegraphics[width=\linewidth]{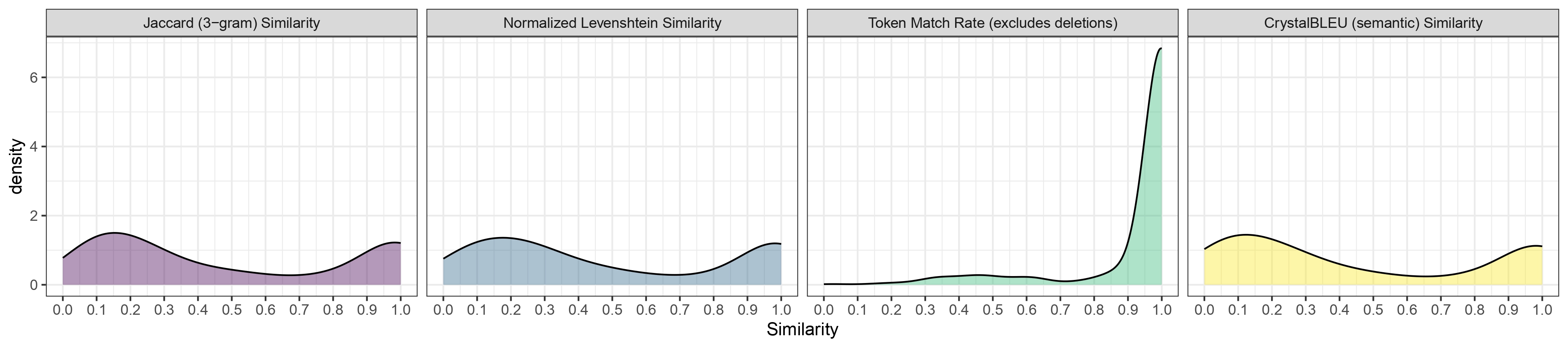}
    \caption{The distribution of similarity scores using Jaccard 3-gram similarity, Normalized Levenshtein similarity, the rate of token matched in the refactored file, and CrystalBLEU similarity (in order).}
    \label{fig:density}
\end{figure*}

\subsection{Adoption of ChatGPT suggestions by developers}
\label{results-RQ1}

While mapping changed files to ChatGPT refactoring suggestions shared in the conversations, we also recorded which prompt (e.g., first, second, etc.) the matched suggestion came from. Note that, due to the nature of our dataset (developers link the ChatGPT conversation in the commit message), there is always an adoption of the refactoring suggestions. Therefore, our results do not cover cases where developers decided not to adopt the suggestion at all.

In Figure \ref{fig:adopted-prompts}, we observe that developers have reached their refactoring goal within the first few prompts (typically, 1-4 prompts). In addition, while there is a clear pattern of using the suggestion that came from the last prompt in the conversation, this did not apply to all conversations. During the manual inspection, we found that this was due to conversations that involve refactoring multiple files, with developers requesting suggestions for different parts of the code at various points. 

After understanding \textit{which} refactoring suggestions were adopted, we look into \textit{how} the suggestions were adopted in terms of the rate of modifications applied to the suggestions. To achieve this, we measure the similarity between ChatGPT's refactoring suggestions and the corresponding final implementation of the developer in 440 committed files using Jaccard 3-gram, Levenshtein, the rate of token match, and CrystalBLEU.

Figure \ref{fig:density} shows the distribution of similarity scores across the four metrics, each represented as a density plot. The density plot shows how similarity scores are distributed by estimating where values are most concentrated, with the area under the curve representing the full dataset of 440 datapoints.
In the first two plots, we see that Jaccard and Levenshtein follow a similar bimodal distribution, where there are two main peaks at the low similarity (0.1-0.3) and high similarity (above 0.9) scores. This indicates that developers either make major modifications (first peak) or adopt the suggestion almost unmodified (second peak), and they less often make minimal changes to suggestions (valley). These modifications can include additions and deletions.

The third (green) plot of Figure \ref{fig:density} shows the distribution of the similarity values using the token match rate. The token match rate checks the proportion of tokens that are in the committed file that also appear in the refactoring suggestion. This means that if the developer deletes parts of the suggestion but uses the remaining parts in the whole file, the similarity score will still be 1.0. We observed that there is more density in higher token match values (above 0.9) with density close to zero in lower values (in other words, the first peak disappears). This shows that the higher rate of modifications and the lower similarity we observed in Jaccard and Levenshtein were mostly due to developers deleting parts of the suggestion and only selecting specific parts to adopt. Hence, the content of most refactored files were adopted from the ChatGPT-generated suggestions. The rate of minor modifications remained consistently low, meaning that these minor modifications are likely to be the developers' own additions or adjustments to the suggestion.

CrystalBLEU in the rightmost plot shows a distribution that is largely consistent with Jaccard and Levenshtein, with two dominant peaks at low and high similarity values. This similarity in distributions can be explained by the fact that CrystalBLEU, although designed to reduce the influence of common or boilerplate tokens, still relies primarily on token overlap between the suggested and final code. Consequently, CrystalBLEU remains sensitive to lexical changes (e.g., deletions or reordering), which results in patterns comparable to those observed in lexical similarity metrics.

\begin{keyfindingsbox}
\textbf{Key finding:} Developers often adopt refactoring suggestions over short interactions (1-4 prompts), then apply them mainly without modifying them or by selecting relevant parts of the suggestion.
\end{keyfindingsbox}

\subsection{Modifications to the suggested code}
\label{results-RQ2}

In our qualitative analysis, we aim to understand the types of refactoring tasks developers commonly seek help with from LLMs, as well as the kinds of modifications they make to the generated suggestions before applying them to their code. To do this, we manually inspected 190 data points, each consisting of the code before the commit, the corresponding ChatGPT suggestion, and the final committed code (see Section \ref{sec:qualitative}).

We present the different refactoring activities that we observed in the 190 datapoints in Table \ref{tab:refactoring_categories}. Note that each datapoint can involve multiple activities. In the conversations, developers often provide specific criteria and refactoring steps that they want ChatGPT to perform (e.g., removing dead code). However, in other cases where developers used general terms such as ``Refactor X'' or ``Enhance Y'', ChatGPT provided the suggestion along with an explanation of what software quality it improved (most commonly, readability and maintainability).

The most common criterion to improve was readability (38\% of the subset) mostly through renaming variables, functions and file names while ensuring that all instances of the old name were correctly updated. Developers also asked ChatGPT to add or modify comments and documentation. In many cases, they asked ChatGPT to clean up the code by removing unnecessary elements such as unused imports, functions, or comments.

Another common goal was improving maintainability (34\%). Developers asked ChatGPT to help restructure their projects and suggest what files should be added or moved. In a few cases, they aimed to adopt object-oriented programming and requested that ChatGPT refactor the code accordingly. Alongside restructuring, ChatGPT often suggested splitting complex logic into smaller and reusable components.

Other use cases focused mainly on simplifying the code by reducing the total lines of code, use more suitable object types or use libraries to replace some parts of their logic with built-in functions.

\begin{table*}[!ht]
\centering
\footnotesize
\caption{Refactoring activities we observed and the specific aspects they aimed to improve.}

\begin{tabularx}{\textwidth}{llX}

\toprule
\textbf{Activities (count)} & \textbf{Improved aspect} & \textbf{Explanation} \\
\midrule
Rename (44) & Readability & Makes names more descriptive. Applies to functions, variables, and file names. \\
Improve documentation (37) & Readability, understandability & Add or extend code documentation (e.g., in-line comments). \\
Restructure (36) & Modularity, maintainability & Improves code organization to support clearer logic and design (e.g., adopting OOP). \\
Splitting logic (33) & Separation of concerns, modularity & Breaks down large elements (e.g., functions) into focused parts. \\
Clean (29) & Maintainability, readability & Removes unused elements (e.g., unused code or imports). \\
Simplify (25) & Reduce complexity & Replaces the logic with simpler or built-in alternatives (e.g., using libraries). \\
Change data type (7) & Maintainability & Replace objects with more suitable types. For example, if a string variable only ever holds two possible values, change it to a boolean variable instead. \\
\bottomrule
\end{tabularx}

\label{tab:refactoring_categories}
\end{table*}


After receiving a code refactoring suggestion, developers can apply some changes before applying it and committing the file. We identified five common patterns that developers follow during the implementation of ChatGPT code refactoring suggestions.

The first pattern is when the \textbf{developer adopts the complete suggestion} and copies it into their file as is without making any modifications, meaning that the similarity scores for all measures were exactly 1.0. However, we noticed that many of these cases are the result of an automated process. In one of the projects, there is a shell script that uses a prompt template to prompt ChatGPT based on the use case (e.g., refactoring), then it extracts the code suggestion, updates the relevant file, and commits the changes. In other cases, the developer provides the entire file(s) as context in their prompt, which results in receiving a refactoring suggestion that can be fully adopted without the need to modify it. The most common refactoring activities that we observed in this pattern were less prone to error (especially when the necessary context is provided), such activities are renaming variables, re-organizing files and improving documentation.

Another pattern is when the suggested parts of the suggested code cause errors or add new behavior such as an additional if statement, error handling, or a new function. In these cases, the \textbf{developer omits the erroneous part or the new behavior}, but adopts the rest of the suggestion. The similarity scores for the cases in this pattern suggest that the changes are rather moderate but are mostly the result of applying deletions rather than additions. The following example shows how a developer removes the additional behavior of copying the response and setting the prompt, which was suggested by ChatGPT. \\

\noindent
\begin{adjustbox}{minipage=\linewidth, margin=0pt 0pt 10pt 0pt}
  \begin{minipage}[t]{0.47\linewidth}
  \footnotesize
  \textbf{ChatGPT Suggestion}
    \scriptsize
  \begin{lstlisting}[style=code, basicstyle=\ttfamily\scriptsize, breaklines=true]
const GenerateButton = () => {
  const handleGeneratePrompt = async () => {
    const response = await generatePrompt();
    (*@\textcolor{codered}{copy(response.prompt)}@*)
      (*@\textcolor{codered}{.then(() => \{ }@*)
        (*@\textcolor{codered}{console.log('Prompt copied to clipboard!');}@*)
      (*@\textcolor{codered}{\})}@*)
      (*@\textcolor{codered}{.catch(err => \{ }@*)
        (*@\textcolor{codered}{console.error('Failed to copy prompt: ', err);}@*)
      (*@\textcolor{codered}{\});}@*)
    (*@\textcolor{codered}{const htmlPrompt = marked(response.prompt);}@*)
    (*@\textcolor{codered}{setPrompt(htmlPrompt);}@*)
  \end{lstlisting}
  \end{minipage}\hfill
  \begin{minipage}[t]{0.47\linewidth}
  \footnotesize
  \textbf{Committed File}
  \begin{lstlisting}[style=code,basicstyle=\ttfamily\scriptsize, breaklines=true]
const GenerateButton = () => {
  const handleGeneratePrompt = async () => {
    const response = await generatePrompt();
  \end{lstlisting}
  \end{minipage}
\end{adjustbox}


We also noted that the conversations can become long due to multiple attempts to fix the refactored code or remove the unwanted behavior using ChatGPT. These interactions appeared to cause frustrations and result in the developer modifying the suggestion independently. For example, in one conversation that consisted of 16 prompts, the developer wanted to refactor multiple files to adopt an object-oriented design, and was continuously providing feedback about the inconsistent logic and the errors caused by the suggestion. The conversation ended with the developer saying: \textit{``no, I'm not doing this because it's ridiculous''}.

The third pattern was mainly caused by the limited access of ChatGPT to the rest of the project, where the \textbf{refactored suggestion is correct in isolation, but incompatible} with the rest of the project. The suggestion relies on certain assumption such as specific function/variable names, file paths, or data types, and sometimes the refactored code relies on a helper function that does not exist. As a result, developers need to make modifications to integrate the suggestion into their code. These modifications are considered minor which are mostly on a token or character level, and occurred only in commits that concern restructuring the code, adopting new design choices (e.g., object-oriented, or the use of functional components), and splitting the logic into different components to increase modularity and ensure the separation of concerns.

Moreover, in another pattern, \textbf{developers applied the suggested simplified logic, but disregard its structure} to better fit their own structural and design choices. For instance, they might combine the logic of multiple suggested functions into one, or split a single function across multiple files. 
Along with these cases, the developers further refactor and improve the code by adding comments and renaming functions to better suit the naming conventions used in the file.
Such structural changes, sometimes combined with additional improvements, typically result in minor or moderate modifications based on Jaccard and Levenshtein similarity scores, both of which are sensitive to token order. However, the token match rate remains high because most of the tokens in the final committed file are still present in the suggestion, just rearranged.

In contrast, other suggestions that scored low similarity to the committed file were due to the \textbf{developer following the structure of the suggestion but disregarding its content}. They would provide a code that has similar functionality to their own, prompt ChatGPT to refactor it, then preserve the layout or function decomposition in the suggestion, while disregarding the logic. On the other hand, sometimes developers apply the suggestions but the similarity remains low due to the nature of the suggestion. For instance, the suggestion may include command line commands to remove unused files, or move other files to new directories, which are not captured by the similarity measures. 
Below is an example of a suggestion to re-organize the files of a project including commands to create and remove directories, and move files:\\



{\footnotesize
\textbf{ChatGPT Suggestion}
\begin{lstlisting}[style=code, basicstyle=\ttfamily\scriptsize, breaklines=true]
  (*@\textcolor{green!40!black}{\# Step 1: Create service/helpers/ directory}@*)
  mkdir -p src/frontend/service/helpers/
  (*@\textcolor{green!40!black}{\# Step 2: Move getComparison to service/helpers/ directory}@*)
  mv src/frontend/services/helpers/getComparison.js src/frontend/service/helpers/getComparison.js
  ...
  (*@\textcolor{green!40!black}{\# Step 5: Delete services/ directory}@*)
  rm -r src/frontend/services/
\end{lstlisting}}

Sometimes, the refactoring suggestion also includes textual instructions, examples, or repeats the content of other files to show the whole context in one code block. Developers then only select and apply the relevant parts of the suggestion to their file.

\begin{keyfindingsbox}
\textbf{Key finding:} 
When adopting refactoring suggestions, developers either make no modifications or minor ones (e.g., change file path) to integrate the refactored suggestion into their code. Larger modifications apply when the suggestion adds new behavior or when developers restructure the suggestion while keeping its core.

\end{keyfindingsbox}


 

%% file: discussion.tex
\section{Discussion } 
\label{discussion}
In this section, we discuss the implications of our results for software practitioners and researchers, as well as validity threats.\\

\textbf{I1: Adoption is inherently complex to measure, but can be reasonably estimated through similarity metrics.}


\citet{russo2024adoption} shows that software engineers and developers are intend to use LLMs mainly to get suggestions on how to refactor and improve their code. However, understanding how developers implement LLM-generated suggestions is hard to study, especially in longer interactions, where multiple suggestions may be used, or where developers may simply be ``inspired'' by a suggestion rather than adopting it directly.
In this study, we analyzed lexical similarity and saw that, overall, the modifications were often substantial. However, in many cases, the suggestions were fully adopted without modification. Then the token match rate revealed that the large modifications were mainly due to deletion operations rather than other changes made by developers. These findings were confirmed later in our manual analysis that resonated with similarity scores consistently. The manual analysis also showed that the actual modifications that the developers make are either minor or in a few cases moderate, otherwise, the changes are mostly due to selecting the relevant parts in the suggestion or deleting unwanted details such as comments and examples. 
This indicates that similarity measures can capture at least one dimension of adoption. However, researchers need to explore additional dimensions (e.g., including multiple suggestions) to develop a more comprehensive measurement and better understand how developers actually use LLM assistance for code-related activities in practice.
\\

\textbf{I2: Developers must ensure LLM suggestions fully qualify as refactoring.}

Refactoring is the process of improving a working code without changing its behavior \cite{fowler2018refactoring}. This means that when developers use LLM-generated refactoring suggestions, they must ensure that (i) the suggested code works, (ii) does not introduce bugs, (iii) does not add new behavior, and (iv) improves code quality. By analyzing the interactions, we observed that checking all the boxes of refactoring in one shot often requires the developer to provide extensive context (e.g., other files in the project) and carefully construct a detailed prompt template. Otherwise, if the LLM is used more as a collaborator, then we saw that ChatGPT could meet the developer's criteria of improving the code quality. However, developers had to make different types of modifications in order to remove additional code elements that introduced bugs or new behaviors, or to ensure that the code works with the rest of the project. This also aligns with previous research showing that LLMs tend to make unnecessary changes to the code while refactoring \cite{shirafuji2023refactoring}, or hallucinate unwanted behavior \cite{cordeiro2024empirical} that may not break the code per se, but would invalidate the refactoring.
\\


\textbf{I3: Automation of refactoring using LLMs is feasible for low-risk tasks.}

The high rates of ``full adoption'' without any modifications that we found through our analysis were mostly due to an automated interaction and adoption of ChatGPT suggestions. These automated processes consist of pre-defined prompt templates that are used to prompt ChatGPT to perform very specific tasks, provide relevant context, and specify how the outcome should look like to enable the automatic extraction and application of the suggestion.
This was interesting because it shows that LLMs are becoming more trusted to automate code-related tasks. However, the types of tasks that were automated were less likely to have high risks of breaking functionality, such as renaming variables and re-organizing files in directories. 
More complex use cases, such as adopting a new software design or splitting the logic into different functions and files, often resulted in the developers making changes before integrating the suggestion. This complements the findings by \citet{cordeiro2024empirical} that found that the LLM refactoring suggestions regarding renaming are more effective compared to complex ones like moving methods that require the developer's attention to the wider context of references and dependencies.
The first step to enable the automation of complex refactoring activities is to have unit and integration tests to ensure that the behavior remains unchanged after refactoring. Since the commits we analyzed lacked tests in the continuous integration (CI) pipeline, ChatGPT's suggestions for major refactorings cannot yet be reliably automated.
\\

\textbf{I4: LLM-generated code suggestions introduce the need for prompt traceability.}

The DevGPT dataset that we used in the study is based on ChatGPT conversations linked in commit messages. This shows that in LLM-assisted software development, there is an emerging need to document the LLM-generated suggestion and the prompt that was used to generate it, and then trace them to the final code in the project. This is particularly important in understanding the rationale behind certain decisions (e.g., merging two functions) and the developer's intent from their prompt.
Furthermore, we noticed during our manual inspection that some developers trace their prompts and conversations in specific files used for documentation (often called \texttt{prompt.md} or \texttt{prompt.yaml}). While it is unclear how developers use the information in these documents, their presence suggests that there is recognition of the value in keeping track of LLM interaction history. This can also present certain tools to accommodate for such needs in software organizations that integrate LLMs in their development process.

\subsection{Threats to validity}
 
\paragraph{Internal Validity}

The manual inspection may introduce some validity threats due to only inspecting a subset rather than the entire dataset, and the possible bias from the authors who conducted the inspection. To address this, the selection strategy that we followed was based on the token match rate in the quantitative results where we covered all 96 instances where there is an indication of a modification (token match rate less than 1.0), then complemented the subset with another 94 datapoints that scored exactly 1.0, which also can include modifications such as deletions. Then, two authors conducted the inspection together to reduce subjectivity and allow them to discuss all modifications to the refactoring suggestions.

\paragraph{External Validity}

The nature of the DevGPT dataset assumes that the developers used at least one suggestion in the ChatGPT conversation they link in the commit message. As a result, it does not capture the cases where developers decided not to adopt a suggestion. Therefore, in this study, we focus mainly on the modifications made by the developers rather than the usefulness of refactoring suggestions to be adopted. 
The dataset is also based on ChatGPT, covering both GPT-3.5 and GPT-4. We did not observe any differences in the results when split based on the model type. However, it is worth investigating other model types and how their performance in code refactoring may impact their rate of adoption.

After processing and extending the dataset, we ended up with 440 datapoints. This subset was unbalanced in terms of the number of GitHub repositories it covers, which risks undermining the representativeness of the different open-source projects and the contributors. Specifically, a concentration of commits from a small number of repositories could bias the observed patterns toward the practices of those projects' contributors. However, we still found consistent patterns across the projects that can be extended with the use of a larger dataset.

\paragraph{Construct Validity}
The refactoring-related keywords that we used involve terms that can be used in non-refactoring context, for example, ``Improve'' that can involve implementing new features. To minimize this threat, we iteratively refined the list of keywords during the experiment and removed some keywords that were producing many false negatives. During the manual analysis, we only found 9 datapoints out of 190 that would not completely qualify as refactoring, which we excluded from the qualitative analysis. We suggest future research to improve the filtering process by using complex keyword search, for instance, ``remov* duplicate*'', to minimize the number of false negatives.

Moreover, we used the lexical similarity to measure the level of adoption. While adoption can be challenging to analyze quantitatively. However, we found the similarity metrics along with the token match rate to resonate with our qualitative observations, which makes lexical similarity measures a reasonable metric to estimate and quantify adoption. However, we believe that further research is still needed to cover all aspects of adoptions.


%% file: conclusion.tex
\section{Conclusion}
In this study, we analyzed developers refactoring commits and corresponding ChatGPT suggestions that they adopted to refactor 440 files. We quantitatively estimate the rate of adoption by measuring the similarity between the suggestion and the final version of the committed code using the rate of matched tokens, Jaccard, Levenshtein similarity and CrystalBLEU. We also qualitatively inspected a subset of 190 datapoints to understand the types of modifications that developers commonly make on different refactoring suggestions.
We found that developers mostly adopt the refactoring suggestions as is, or select only parts of the suggestions to adopt with occasional minor changes to integrate the suggestion into their code (e.g., updating file paths). We also noticed that the cases of full adoption were mostly on less complex refactoring tasks such as renaming and reorganizing files in a project. Many of these were also the result of an automated process of prompting ChatGPT and adopting the suggestion.
In other cases with more complex refactoring activities, such as adopting a new software design, ChatGPT failed to provide a valid refactoring suggestion. In particular, the suggestion often caused errors or added new behavior to the code, which required the developers to make modifications of different levels (e.g., fixing data type to re-writing functions).
Our work can be extended in future research to see how our findings generalize to other open-source projects and projects in software organizations that adopted an AI-assisted development process. In addition, more work is needed to explore different aspects of adoption that need to be captured in quantitative and qualitative measures.